\documentclass[sigconf]{acmart}
\acmArticleType{Review}
\AtBeginDocument{%
  }

\setcopyright{acmlicensed}
\copyrightyear{2018}
\acmYear{2018}
\acmDOI{XXXXXXX.XXXXXXX}
\acmISBN{978-1-4503-XXXX-X/2018/06}

\usepackage{multirow}
\usepackage[table]{xcolor}
\usepackage{booktabs}
\usepackage{subcaption}
\usepackage{enumitem}
\usepackage{algorithm}
\usepackage{algpseudocode}

\newcommand{\ie}{\emph{i.e., }}

\definecolor{blue-violet1}{rgb}{0.20, 0.20, 0.60}
\newcommand{\dhx}[1]{\textcolor{black}{#1}}
\definecolor{blue-violet}{rgb}{0.54, 0.17, 0.89}
\newcommand{\zjz}[1]{\textcolor{black}{#1}}

\begin{document}

\title{Towards Sample-Efficient and Stable Reinforcement Learning for LLM-based Recommendation}






\author{Hongxun Ding, Keqin Bao, Jizhi Zhang, Yi Fang, Wenxin Xu, Fuli Feng, Xiangnan He}
\authornote{Corresponding author.}
\affiliation{
  \institution{University of Science and Technology of China}
  \country{China}
}

\email{hongxunding02@gmail.com, {baokq, cdzhangjizhi, peterfang, xuwenxin}@mail.ustc.edu.cn}
\email{{fulifeng93, xiangnanhe}@gmail.com}
\renewcommand{\shortauthors}{Ding et al.}

\begin{abstract}

While Long Chain-of-Thought (Long CoT) reasoning has shown promise in Large Language Models (LLMs), its adoption for enhancing recommendation quality is growing rapidly. In this work, we critically examine this trend and argue that Long CoT is inherently ill-suited for the sequential recommendation domain. We attribute this misalignment to two primary factors: excessive inference latency and the lack of explicit cognitive reasoning patterns in user behavioral data. Driven by these observations, we propose pivoting away from the CoT structure to directly leverage its underlying mechanism: Reinforcement Learning (RL), to explore the item space. However, applying RL directly faces significant obstacles, notably low sample efficiency—where most actions fail to provide learning signals—and training instability. To overcome these limitations, we propose \textbf{RISER}, a novel \textbf{R}einforced \textbf{I}tem \textbf{S}pace \textbf{E}xploration framework for \textbf{R}ecommendation. RISER is designed to transform non-learnable trajectories into effective pairwise preference data for optimization. Furthermore, it incorporates specific strategies to ensure stability, including the prevention of redundant rollouts and the constraint of token-level update magnitudes. Extensive experiments on three real-world datasets show that RISER significantly outperforms competitive baselines, establishing a robust paradigm for RL-enhanced LLM recommendation. Our code will be available at \url{https://anonymous.4open.science/r/RISER/}.
\end{abstract}


\received{20 February 2007}
\received[revised]{12 March 2009}
\received[accepted]{5 June 2009}

\maketitle

\section{Introduction}



In recent years, Large Language Models (LLMs) have advanced the frontier of recommender systems. 
Researchers have leveraged their powerful semantic understanding and generation capabilities to model user behaviors and item attributes, aiming to deliver more accurate recommendations~\cite{ bao2023tallrec, bao2025bi, gao2023chat,sun2024large, zhang2024agentcf,wang2024recmind,zhang2024text,bao2024decoding,sheng2025language}. 
Meanwhile, driven by the emergence of Reasoning Large Language Models (RLLMs) such as OpenAI o1~\cite{jaech2024openai} and DeepSeek R1~\cite{guo2025deepseek}, Long Chain-of-Thought (Long CoT) reasoning has garnered significant attention~\cite{chen2025towards, sun2023survey, zhong2024evaluation}. 
The core idea of Long CoT is to simulate human thought processes by generating intermediate reasoning steps at test time. Essentially, it's a strategy that trades increased computational overhead for extended ``thinking'' time, improving the reliability of final results. 
In recommender systems, Long CoT can generate reasoning paths to infer users' complex and dynamic preferences, thereby enabling more accurate and interpretable recommendations. Consequently, a growing body of work~\cite{bismay2024reasoningrec, fang2025reason4rec, zhang2025reinforced, zhao2025reason, xie2025recllm} has leveraged \dhx{reasoning-enhanced approaches to improve recommendation performance using LLMs}.


However, this promising trend overlooks two foundational challenges that question the direct applicability of Long CoT in \dhx{Sequential Recommendation (SeqRec)}. The first is \textbf{Inference Latency}. The generation of extended reasoning chains introduces substantial computational overhead, leading to inference latencies that are often tens or even hundreds of times greater. This severe performance degradation is \dhx{fundamentally aligned against} the low-latency requirements of real-world recommender systems, creating a significant barrier to practical deployment.
More critically, the second challenge is \textbf{Cognitive Pattern Scarcity}. The efficacy of Long CoT, particularly when enhanced via Reinforcement Learning (RL), hinges not on correct final answers but on the presence of identifiable cognitive patterns—such as ``verification'' or ``backtracking''—within the pre-training data~\cite{gandhi2025cognitive}. Domains like mathematics and coding are rich with such explicit reasoning signals. \dhx{SeqRec data}, in contrast, is inherently behavioral, recording \dhx{only implicit interaction sequences} (e.g., clicks, purchases) while leaving the underlying thought processes entirely unobserved. \dhx{This creates a structural mismatch: the implicit, sparse signals of ID-based sequences fail to trigger the explicit reasoning patterns acquired during pre-training, thereby hindering their adaptation into domain-specific capabilities. The} disconnect represents a core bottleneck, impeding the effective transfer of Long CoT's success from reasoning-intensive tasks to \dhx{SeqRec}.


\dhx{The aforementioned challenges reveal a fundamental misalignment between the Long CoT framework and the intrinsic characteristics of SeqRec.} This compels us to dissect its underlying mechanism, and we argue that the true power of Long CoT stems from the RL paradigm that enables active exploration. Therefore, we propose to \dhx{bypass} the ill-suited structure of Long CoT and instead apply RL directly to recommendation tasks. The crucial advantage of RL over Supervised Fine-Tuning (SFT) is its inherent capacity to explore the item space via policy-guided rollouts. This allows the model to transcend the limitations of merely imitating historical user-item interactions, thereby mitigating the imitation bias inherent in SFT (e.g., popularity bias) and autonomously correcting recommendation inaccuracies. To this end, inspired by DeepSeek R1~\cite{guo2025deepseek}, we devise a two-stage training framework consisting of a SFT stage before the RL stage that employs the Group Relative Policy Optimization (GRPO)~\cite{shao2024deepseekmath} algorithm.

However, directly applying this method presents two main challenges:
(a) \textbf{Low Sample Utilization}: 
The model learns inefficiently because most of its attempts are unsuccessful.\dhx{\footnote{\dhx{Empirically, around 80-90\% of samples yield no correct hit during early rollouts.}}}
These failed attempts offer no useful feedback, meaning the model doesn't learn from them. This makes the training process very wasteful.
(b) \textbf{Training Instability}: The model can become unstable during training for two reasons. First, it often repeatedly generates the same items during RL rollouts. This limits its ability to discover a diverse range of better items. Second, the way item IDs are represented as text can cause problems~\cite{bao2024decoding}. When a large number of items share a common prefix (e.g., ``BrandX-A-...''), the model during RL training may over-learn to this pattern, suppressing its ability to explore items with similar prefixes (e.g., ``BrandX-B-...'') and consequently degrading its recommendation capability. This leads to an unstable and inefficient learning process. Besides, for highly predictable, deterministic tokens, aggressive updates are both unnecessary and inefficient, yet the current training paradigm fails to differentiate between these cases.

To address the challenges outlined above, we introduce \textbf{RISER} (\textit{\underline{R}einforced \underline{I}tem \underline{S}pace \underline{E}xploration for \underline{R}ecommendation}), a novel framework for end-to-end policy learning in recommendation.
To enhance sample utilization, RISER incorporates Simple Preference Optimization (SimPO)~\cite{meng2024simpo}, which repurposes zero-advantage trajectories into preference data, thereby ensuring that learning signals are extracted from all rollouts.
To improve training stability, we introduce several targeted mechanisms. We first employ oversampling and de-duplication to increase rollout diversity and mitigate the model's tendency for repetitive sampling. Additionally, we stabilize the token-level learning dynamics using two techniques: a KL-Cov~\cite{cui2025entropy} regularization strategy selectively penalizes outlier tokens (\ie those with high confidence and advantage) to ensure smooth convergence, and a loss mask down-weights updates for highly predictable tokens. Inspired by prior work~\cite{bao2024decoding}, we also remove the length penalty, which can destabilize training. We conduct extensive experiments on three real-world datasets to validate RISER's capabilities and provide further analysis.

In conclusion, our contributions are as follows: 
\vspace{-0.09cm}
\begin{itemize}[leftmargin=*]

\item We identify a fundamental \dhx{misalignment} between Long CoT and \dhx{SeqRec} tasks, arising from prohibitive inference costs and a critical mismatch in cognitive patterns. This analysis redirects focus toward the core exploratory power of RL.
\item  We propose RISER, a novel RL framework for recommendation specifically designed to surmount two critical barriers to its direct application: low sample inefficiency and training instability.
\item  Comprehensive experiments on three real-world datasets demonstrate that RISER significantly outperforms strong baselines, validating its efficacy and robustness.
\end{itemize}
\vspace{-0.2cm}
\section{Preliminaries}

We frame SeqRec as a conditional language generation problem. Our training protocol follows a two-stage paradigm: it first involves SFT to learn the recommendation task's fundamentals, followed by RL for policy optimization to further enhance recommendation performance.

\subsection{Problem Formulation and Dataset}
Given a dataset $\mathcal{D} = \{(u, h, y)\}$, where $u$ denotes a user, $h$ is their chronological sequence of interacted items, and $y$ is the next item, the goal is to predict $y$. All items are represented by their natural language descriptions (e.g., item titles). Each data instance is transformed into an input-output pair $(x, y)$, where $x$ is a structured textual prompt derived from the user's context $(u, h)$. Detailed prompt structure is available in Appendix~\ref{sec:prompt}.

The SFT stage uses the training set $\mathcal{D}_{\text{SFT}}$, while the RL stage employs a smaller training set $\mathcal{D}_{\text{RL}}$. This two-stage data approach serves two purposes. First, the smaller set $\mathcal{D}_{\text{RL}}$ is used for efficiency, as the RL stage is far more computationally intensive than SFT. Second, and more importantly, using data for RL that was unseen during SFT promotes generalization. It forces the policy to learn a robust strategy rather than merely exploiting patterns it may have memorized from $\mathcal{D}_{\text{SFT}}$.

\subsection{Supervised Fine-Tuning}
Applying RL directly to a general-purpose LLM for recommendation is infeasible. Given the recommendation task's vast action space and sparse reward signals, a naive policy can hardly obtain any positive feedback through random exploration, leading to training failure.
To address this cold start problem, we first employ SFT. The model is trained on the formatted $(x, y)$ pairs from the $\mathcal{D}_{\text{SFT}}$ set with the objective of minimizing the expected negative log-likelihood:
\begin{equation}
    \mathcal{L}_{\text{SFT}} = - \mathbb{E}_{(x,y) \sim \mathcal{D}_{\text{SFT}}} [\log p(y|x)].
    \label{eq:sft_objective}
\end{equation}
This SFT stage serves as a crucial warm-up. It aligns the general LLM with the recommendation domain, providing a solid policy initialization where the model can already generate plausible and contextually relevant items. This makes the subsequent, more expensive exploration in the RL stage both tractable and effective.

\subsection{Reinforcement Learning}
A subsequent RL stage is employed to enhance the model's ability to generate high-quality recommendations beyond simply mimicking training data. RL complements the SFT stage by allowing the model to learn directly from reward signal, facilitating continuous self-improvement. For this stage, the default choice is GRPO algorithm, which stabilizes policy updates by computing group-normalized advantages over multiple sampled completions. The GRPO objective, which is maximized, is defined as:
\begin{equation}
\begin{split}
J_{\text{GRPO}}(\theta)=
& \mathbb{E}_{q, \{o_i\}}\Bigg[ \frac{1}{G} \sum_{i=1}^{G} \min\Big(r_i(\theta) A_i, \text{clip}(r_i(\theta), 1-\epsilon, 1+\epsilon)A_i\Big) \Bigg] \\
&- \beta_{\text{KL}} D_{\text{KL}}(\pi_\theta \| \pi_{\text{ref}}).
\end{split}
\label{eq:grpo_objective}
\end{equation}
Here, $J_{\text{GRPO}}(\theta)$ is the objective function parameterized by the policy's weights $\theta$. The expectation $\mathbb{E}_{q, \{o_i\}}$ is taken over queries $q$ and a group of $G$ completions (i.e., generated items) $\{o_i\}_{i=1}^G$ sampled for each query. The core of the objective relies on the probability ratio $r_i(\theta) = \frac{\pi_\theta(o_i|q)}{\pi_{\theta_{\text{old}}}(o_i|q)}$, which compares the likelihood of generating a given completion between the current policy $\pi_\theta$ and the old policy $\pi_{\theta_{\text{old}}}$ used for sampling. This ratio is then multiplied by the group-relative advantage $A_i$ for that completion. To ensure stability, a $\text{clip}(\cdot)$ function constrains this ratio to the range $[1-\epsilon, 1+\epsilon]$, where $\epsilon$ is a clipping hyperparameter. The final term provides regularization, where $D_{\text{KL}}(\pi_\theta \| \pi_{\text{ref}})$ is the Kullback–Leibler divergence between the current policy and a reference policy $\pi_{\text{ref}}$ (\ie the SFT model), weighted by the coefficient $\beta_{\text{KL}}$. In our protocol, the reward function used to calculate the advantages is direct: a positive reward for a ``hit'' and a negative reward otherwise.

However, directly applying this GRPO objective to recommendation reveals two key limitations.
\begin{itemize}[leftmargin=*]
\item \textbf{Low Sample Utilization}: While sparse rewards are a general RL problem, this issue is acutely magnified in recommendation due to the limited exploratory capacity of the SFT-trained policy. Although SFT aligns the model with the task, its policy distribution is often too narrow to sample correct but lower-probability items during exploratory rollouts. Consequently, the vast majority of these trajectories fail to hit any correct items, which renders the group-relative advantage $A_i$ and the policy gradient $\nabla_\theta J$ null. This low sample utilization makes the learning process extremely inefficient, preventing the policy from effectively improving.
\item \textbf{Training Instability}: The second challenge arises from the unique features of the recommendation item space, which the standard GRPO protocol is ill-equipped to handle. On one hand, the RL stage is susceptible to repetitive rollouts generation. 
When the standard GRPO objective is fed this stream of repetitive rollouts, it causes the policy to collapse onto a few items and stifling exploration. 
On the other hand, the token-based structure of item IDs creates fine-grained training instabilities.
This leads to two distinct issues. First, it causes disproportionate updates on distinguishing tokens. Since similar items often differ by only a few critical tokens, such as ``BrandX-\textbf{A}-...'' and ``BrandX-\textbf{B}-..'', any imbalance in the training batch can cause the gradient signal for those specific tokens to become excessively large, destabilizing the output distribution and suppressing the exploration of valid alternatives in subsequent learning steps, thereby hindering effective convergence.
Second, for highly predictable, deterministic tokens, aggressive updates are both unnecessary and inefficient, yet the current training paradigm fails to differentiate between these cases. The global $D_{\text{KL}}$ regularization term is too coarse to address these specific, token-level problems.
\end{itemize}



\section{Methods}

\begin{figure*}[htbp]
    \centering
    \includegraphics[width=1.85\columnwidth]{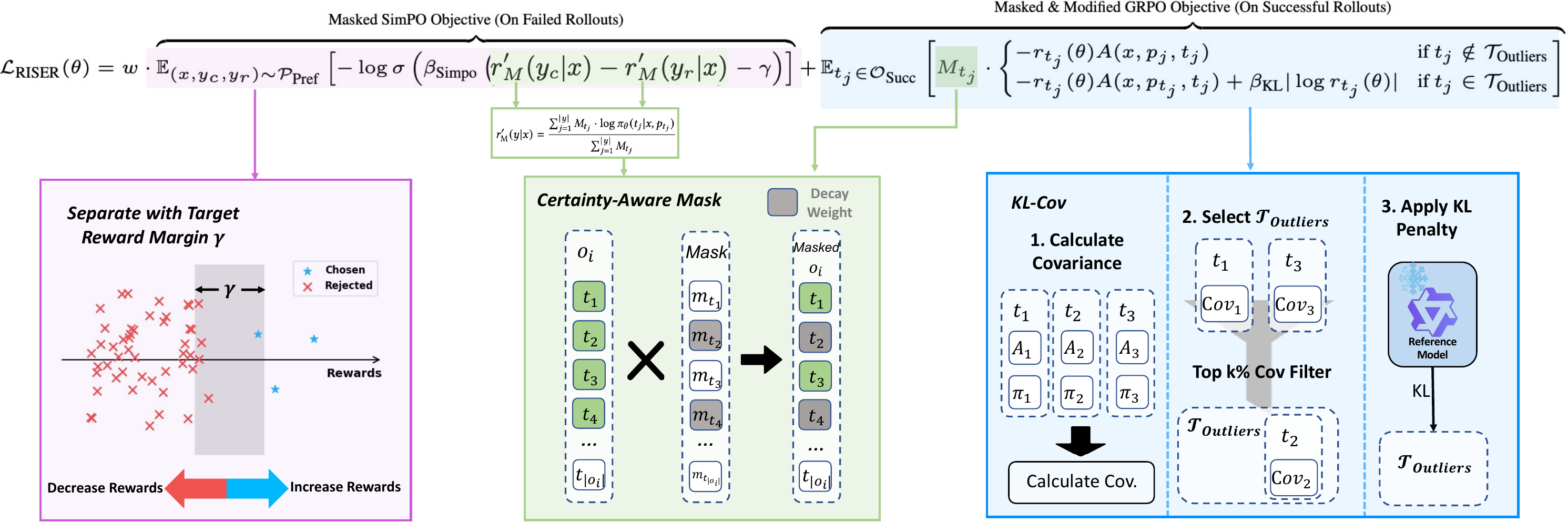}
    \caption{Overview of the RISER framework. The final loss, $\mathcal{L}_{\text{RISER}}$ is a weighted sum of two objectives applied to different data partitions: a modified GRPO objective for successful rollouts and a SimPO objective for failed rollouts. Each core mechanism is detailed in the corresponding sub-diagram.}
    \label{fig:main_figure_draft}
\end{figure*}

To address the core challenges of inefficient sample utilization and training instability that arise from the direct application of GRPO to recommendation, we propose the \textbf{RISER} (\textit{\underline{R}einforced \underline{I}tem \underline{S}pace \underline{E}xploration for \underline{R}ecommendation}) framework. RISER refines the SFT-GRPO pipeline, enabling robust and efficient end-to-end policy learning directly in the item space. Our design is composed of two primary modules that tackle each challenge respectively. An overview of the RISER framework is illustrated in Figure~\ref{fig:main_figure_draft}.

\subsection{Enhancing Sample Utilization}
A primary bottleneck in applying GRPO is the low sample utilization from zero-advantage trajectories. To address this, RISER employs the SimPO~\cite{meng2024simpo} algorithm to extract a valuable learning signal from these wasted rollouts.

Specifically, we first identify training instances where all $G$ rollouts for a prompt $x$ fail to generate the ground-truth item $y$. For each instance, we construct $G$ preference pairs. In each pair, the preferred (chosen) response, $y_c$, is the ground-truth item $y$, while the dispreferred (rejected) response, $y_r$, is one of the incorrectly generated items.
SimPO models the preference probability based on the Bradley-Terry objective~\cite{bradley1952rank}. It defines a base reward for any completion $y$ as its average log-likelihood, $r'(y|x) = \frac{1}{|y|} \log \pi_\theta(y|x)$. The final SimPO loss is derived by taking the negative log-likelihood over the constructed set of preference pairs, $\mathcal{P}_{\text{Pref}}$:
\begin{equation}
    \mathcal{L}_{\text{SimPO}} = -\mathbb{E}_{(x, y_c, y_r) \sim \mathcal{P}_{\text{Pref}}} \log \sigma \left( \beta_{SimPO} \left( r'(y_c|x) - r'(y_r|x) \right) - \gamma \right).
    \label{eq:simpo_loss_final}
\end{equation}
Here, $\beta_{Simpo}$ is a constant that scales the reward difference, controlling the sharpness of the preference. $\gamma$ is a constant representing the target reward margin, which enforces a minimum gap between the scaled rewards of preferred and dispreferred responses. By converting all zero-advantage rollouts into a rich set of preference data, SimPO directly resolves the sample inefficiency bottleneck.



\subsection{Improving Training Stability}
The second major challenge is maintaining training stability during RL. RISER introduces a suite of three complementary techniques, specifically designed for LLM-based recommendation, to ensure a stable and effective RL learning process.

\subsubsection{Diversifying Rollouts via Oversampling and De-duplication}
The SFT-trained policy's tendency to generate repetitive rollouts creates two distinct problems for the RL phase: it stifles exploration by limiting the variety of sampled items, and it creates imbalanced training batches that can lead to disproportionate gradient updates on certain distinguishing tokens.
To encourage the model to explore new items and address both issues above, we introduce an oversampling and de-duplication strategy. Formally, let $n$ be the target number of completions required for the GRPO update. 
Instead of generating $n$ rollouts, we oversample a larger multiset \(O_{\text{Multi}}\) of m completions (m>n), where each completion \(o_i\) is drawn independently from the policy \(\pi_{\theta_{\text{old}}}\).
To balance the need for diversity against the computational cost of generation, $m$ is set to be only moderately larger than $n$:
\begin{equation}
    O_{\text{Multi}} = \{o_1, o_2, \dots, o_m\}, \quad o_i \sim \pi_{\theta_{\text{old}}}(o|x).
\end{equation}
From this multiset, we first extract the set of unique completions, $O_{\text{Unique}}$, and the multiset of duplicates, $O_{\text{Dups}} = O_{\text{Multi}} \setminus O_{\text{Unique}}$. We then construct the final set of $n$ rollouts, $O_{\text{Final}}$, for the GRPO update:
\begin{equation}
O_{\text{Final}} =
\begin{cases}
    \text{sample}(O_{\text{Unique}}, n) & \text{if } |O_{\text{Unique}}| \geq n \\
    O_{\text{Unique}} \cup \text{sample}(O_{\text{Dups}}, n - |O_{\text{Unique}}|) & \text{if } |O_{\text{Unique}}| < n,
\end{cases}
\label{eq:oversampling}
\end{equation}
where the sample$(\mathcal{S}, k)$ function denotes the operation of randomly selecting a subset of k elements from the set $\mathcal{S}$ without replacement. This procedure alleviates the repetition in the batch fed to GRPO, encouraging broader exploration and stabilizing training process.

\subsubsection{Focusing Gradients with a Certainty-Aware Mask}
One of the source causing training instability is the inefficient updates on highly predictable tokens that arise from the standard GRPO objective. Because the objective applies updates indiscriminately to all tokens, frequent updates are imposed on the predictable, non-branching parts of an item's ID sequence.  To solve this, we introduce a certainty-aware loss mask, which strategically down-weights the loss for these deterministic tokens, thereby concentrating the learning signal on the critical branching points and stabilizing the training process.

\zjz{To address this, first, we pre-process the entire item vocabulary to build a prefix tree, denoted by $\mathcal{T}_{Item}$. This structure allows us to know, for any given token prefix, how many valid next tokens exist. Let $p_{t_j} = (t_1, \dots, t_{j-1})$ be a prefix of token \(t_j\) for a generated item, and let $\text{Children}(p_{t_{j}})$ be the set of all possible valid next tokens following this prefix in $\mathcal{T}_{Item}$.}

\zjz{A token $t_{j}$ is considered a decisive, non-deterministic choice if its preceding prefix $p_{t_{j}}$ is a branching point in the tree, i.e., if $|\text{Children}(p_{t_{j}})| > 1$. Conversely, if $|\text{Children}(p_{t_{j}})| = 1$, the token $t_{j}$ is part of a deterministic path.}

Based on this, we construct a mask $M$ for each generated sequence $o = \zjz{(t_1, \dots, t_{|o|})}$. The weight for each token $t_j$ in the sequence is defined as:

\begin{equation}
M_{t_{j}} =
\begin{cases}
    1 & \text{if } |\text{Children}(p_{t_{j}})| > 1 \text{ or } t_j \text{ is the final token} \\
    d & \text{otherwise},
\end{cases}
\label{eq:certainty_mask}
\end{equation}
where $p_{t_{j}}$ is the prefix preceding token $t_{j}$, and $d \in [0, 1)$ is a decay coefficient that down-weights updates on non-branching, deterministic tokens. 

This mask $M$ is then integrated directly into the loss calculations for both RL objectives. For the GRPO objective, it is applied as an element-wise weight to the entire token-level loss computation, which includes both the policy gradient and the selective KL-Cov penalty terms described in the next section. For the SimPO objective, the mask is applied when computing the base reward $r'(y|x)$, transforming it from a simple average of log-probabilities into a weighted average:
\zjz{\begin{equation}
    r'_{\text{M}}(y|x) = \frac{\sum_{j=1}^{|y|} M_{t_{j}} \cdot \log \pi_\theta(t_{j}|x, p_{t_{j}})}{\sum_{j=1}^{|y|} M_{t_j}}.
    \label{eq:masked_reward}
\end{equation}}
By applying the mask in this manner, we ensure that the model could focus its updates on the truly decisive parts of an item's representation.

\dhx{
We implement the prefix tree $\mathcal{T}_{Item}$ using a Hash Map structure, which offers two distinct advantages:
(1) \textbf{Scalability:} It supports incremental updates for cold-start items without rebuilding the entire tree. New items are inserted in $O(l)$ time (where $l$ is the length of the tokenized item ID) by simply updating the hash entries.
(2) \textbf{Space Efficiency:} The structure naturally compresses storage by sharing common prefixes across items, ensuring low memory usage for large-scale item sets.
}

\subsubsection{Stabilizing Entropy with KL-Cov}

The standard GRPO objective includes global KL regularization term $D_{\text{KL}}$ intended for stabilization. However, it fails to resolve the issue of gradient over-updating on distinguishing tokens and is ineffective at preventing policy collapse, as its uniform suppression of exploration can degrade policy performance~\cite{he2025skywork}.

To provide a more precise and effective solution, we replace this global regularizer with KL-Cov~\cite{cui2025entropy}.
KL-Cov achieves this by applying the penalty only to a small subset of outlier tokens, which are identified dynamically in each training batch based on their statistical performance. An outlier token is one where the policy is both highly confident and effective. This mechanism could mitigate over-updating issue on distinguishing tokens. For instance, if a high-reward item is over-represented in a batch, its distinguishing tokens will likely exhibit both higher-than-average log-probability and advantage. These tokens are thus flagged as outliers, and the subsequent KL penalty tempers their updates, preventing their probabilities from being excessively amplified and stopping the policy from collapsing onto a single, dominant item.
Formally, this is implemented in three steps. First, we compute a token-wise covariance score on successful trajectories to identify the outliers. For each token \zjz{$t_j$}, the score is:
\begin{equation}
\small
    \text{Cov}(t_{j}) = (\log \pi_\theta(t_{j}|x, p_{t_{j}}) - \mathbb{E}_{t}[\log \pi_\theta]) \cdot (A(x, p_{t_j}, t_j) - \mathbb{E}_t[A]).
    \label{eq:kl_cov}
\end{equation}
Here, \zjz{$\mathbb{E}_{t}[\cdot]$} denotes the mean over all valid tokens in the batch, and the per-token advantage \zjz{$A(x, p_{t_j}, t_j)$} is derived from the sequence-level advantage. Second, we select a top-$k$ proportion of tokens with the highest scores to form the set $\mathcal{T}_{\text{Outliers}}$. Finally, the modified GRPO loss, $\mathcal{L}'_{\text{GRPO}}$, is defined using the per-token probability ratio, 
\zjz{$r_{t_j}(\theta) = \frac{\pi_\theta(t
_j|x, p_{t_{j}})}{\pi_{\theta_{\text{old}}}(t_j|x, p_{t_{j}})}$}, 
to apply the penalty only to these outliers:

\zjz{\begin{equation}
\small
\mathcal{L}'_{\text{GRPO}}(t_j; \theta) =
\begin{cases}
    -r_{t_{j}}(\theta) A(x, p_{j}, t_j) & \text{if } t_j \notin \mathcal{T}_{\text{Outliers}} \\
    -r_{t_j}(\theta) A(x,  p_{t_{j}}, t_j) + \beta_{\text{KL}} |\log r_{t_j}(\theta)| & \text{if } t_j \in \mathcal{T}_{\text{Outliers}}.
\end{cases}
\label{eq:grpo_prime_loss}
\end{equation}}
This entire token-level loss is then weighted by the certainty-aware mask (Eq.~\ref{eq:certainty_mask}) before the final expectation is taken.
This targeted regularization resolves the stability-exploration trade-off more effectively than a global KL penalty, making the training process more stable and ultimately enhancing recommendation performance.

\subsection{Final Training Algorithm}
The complete RISER training process consists of two sequential stages. The first stage is SFT. The second, iterative RL stage employs an on-policy update mechanism. Within each iteration, a set of rollouts, $\mathcal{O}_{\text{Rollout}}$, is generated and then partitioned based on performance. Trajectories containing at least one ground-truth item form the successful set $\mathcal{O}_{\text{Succ}}$, while those with none form the failed set $\mathcal{O}_{\text{Fail}}$. The modified GRPO loss is computed on $\mathcal{O}_{\text{Succ}}$, while a set of preference pairs, $\mathcal{P}_{\text{Pref}}$, is constructed from $\mathcal{O}_{\text{Fail}}$ to compute the SimPO loss. These two losses are then combined to update the policy, as shown in Figure ~\ref{fig:main_figure_draft}.
The detailed pseudo code can be found in Appendix~\ref{sec:algo}.

\dhx{
\noindent\textbf{Complexity Analysis.} 
We compare the efficiency of RISER against Long CoT methods. Let $L_{\text{Item}}$ and $L_{\text{CoT}}$ denote the average token lengths of item sequences and reasoning chains, respectively.
Since RISER bypasses intermediate reasoning, its time and KV-cache space complexity scales as $\mathcal{O}(L_{\text{Item}})$.
In contrast, Long CoT methods scale as $\mathcal{O}(L_{\text{CoT}} + L_{\text{Item}})$.
Given that typically $L_{\text{CoT}} \gg L_{\text{Item}}$, RISER significantly reduces the computational burden, offering superior efficiency in latency and memory usage.
}

\section{Experiments}
In this section, we conduct extensive experiments to answer the following research questions:

\begin{itemize}[leftmargin=*]
\item {\textbf{RQ1}}: How does RISER's recommendation effectiveness compare against conventional and LLM-based methods?
\item {\textbf{RQ2}}: What are the primary factors contributing to the performance improvements of RISER? 
\item {\textbf{RQ3}}: How effectively does RISER maintain training stability, particularly in mitigating bias and preventing entropy collapse during RL?
\item {\textbf{RQ4}}: How does the performance of RISER scale with an increasing volume of training data?
\end{itemize}
\subsection{Experimental Settings}

\subsubsection{Datasets.} 
We conduct experiments on three public, real-world datasets spanning two distinct domains: e-commerce (Amazon Games and Toys~\cite{ni2019justifying}) and media consumption (Goodreads~\cite{dblp:conf/acl/wanmnm19,dblp:conf/recsys/wanm18}). This diverse set of benchmarks enables us to evaluate the general effectiveness of our model across various contexts. Following previous works~\cite{bao2025bi,kang2018sasrec}, the raw data undergoes several preprocessing steps. First, we employ a 5-core filtering protocol, removing users and items with fewer than five interactions. Second, all interaction sequences are truncated to a maximum length of 10. Third, to ensure a proper dataset size, a custom sampling method is applied to each dataset to yield a final item count exceeding 10,000. 
For each dataset, we partition the interactions chronologically into training, validation, and test sets with a ratio of 8:1:1.
This chronological partitioning simulates a realistic scenario and ensures that evaluation is performed on future, unseen data, thus preventing information leakage~\cite{ji2023critical}. Detailed statistics and the full preprocessing procedure are provided in Appendix~\ref{sec:dataset_statistics}.
\begin{table*}[t]
\centering
\setlength{\tabcolsep}{1.5pt} 
\caption{\dhx{Top-N recommendation performance comparison of our method (RISER) and various baselines on three distinct datasets. N and H denote NDCG and HR, respectively. The best results are highlighted in \textbf{bold}, and the second-best results are \underline{underlined}. ``Improv.'' denotes the relative improvement of RISER over the best-performing baseline (D$^{3}$) for each metric.}}
\resizebox{\textwidth}{!}{%
\begin{tabular}{c|l|ccc|ccc|ccccc|cc}
\toprule
Datasets & Metrics & Caser & GRU4Rec & SASRec & Caser$^{*}$ & GRU4Rec$^{*}$ & SASRec$^{*}$ & AlphaRec & D$^{3}$ & S-DPO & CoT-RL & LatentR$^3$ & \textbf{Ours} & \cellcolor{gray!20}Improv. \\
\midrule
\multirow{6}{*}{\textbf{Games}}
& N@5  & 0.0163 & 0.0161 & 0.0247 & 0.0173 & 0.0329 & 0.0360 & 0.0294 & 0.0378 & 0.0322 & 0.0074 & \underline{0.0387} & \textbf{0.0526} & \cellcolor{gray!20}+35.92\% \\
& N@10 & 0.0217 & 0.0221 & 0.0313 & 0.0216 & 0.0401 & 0.0417 & 0.0378 & \underline{0.0447} & 0.0388 & 0.0097 & 0.0439 & \textbf{0.0598} & \cellcolor{gray!20}+33.78\% \\
& N@20 & 0.0270 & 0.0287 & 0.0391 & 0.0265 & 0.0477 & 0.0476 & 0.0479 & \underline{0.0516} & 0.0461 & 0.0126 & \underline{0.0516} & \textbf{0.0677} & \cellcolor{gray!20}+31.20\% \\
& H@5  & 0.0244 & 0.0242 & 0.0354 & 0.0260 & 0.0460 & 0.0482 & 0.0446 & 0.0502 & 0.0444 & 0.0108 & \underline{0.0536} & \textbf{0.0680} & \cellcolor{gray!20}+26.87\% \\
& H@10 & 0.0412 & 0.0428 & 0.0558 & 0.0396 & 0.0684 & 0.0660 & 0.0710 & \underline{0.0714} & 0.0648 & 0.0172 & 0.0698 & \textbf{0.0902} & \cellcolor{gray!20}+26.33\% \\
& H@20 & 0.0626 & 0.0690 & 0.0870 & 0.0594 & 0.0988 & 0.0894 & \underline{0.1110} & 0.0986 & 0.0936 & 0.0276 & 0.1002 & \textbf{0.1214} & \cellcolor{gray!20}+9.37\% \\
\midrule
\multirow{6}{*}{\textbf{Toys}}
& N@5  & 0.0221 & 0.0196 & 0.0307 & 0.0193 & 0.0482 & 0.0455 & 0.0265 & 0.0577 & \underline{0.0605} & 0.0227 & 0.0529 & \textbf{0.0660} & \cellcolor{gray!20}+9.09\% \\
& N@10 & 0.0266 & 0.0234 & 0.0349 & 0.0251 & 0.0516 & 0.0497 & 0.0353 & 0.0642 & \underline{0.0671} & 0.0308 & 0.0604 & \textbf{0.0734} & \cellcolor{gray!20}+9.39\% \\
& N@20 & 0.0307 & 0.0272 & 0.0389 & 0.0295 & 0.0563 & 0.0528 & 0.0445 & 0.0693 & \underline{0.0736} & 0.0371 & 0.0645 & \textbf{0.0794} & \cellcolor{gray!20}+7.88\% \\
& H@5  & 0.0316 & 0.0264 & 0.0408 & 0.0288 & 0.0606 & 0.0584 & 0.0420 & 0.0744 & \underline{0.0790} & 0.0325 & 0.0714 & \textbf{0.0842} & \cellcolor{gray!20}+6.58\% \\
& H@10 & 0.0456 & 0.0384 & 0.0538 & 0.0470 & 0.0714 & 0.0712 & 0.0694 & 0.0946 & \underline{0.0994} & 0.0506 & 0.0946 & \textbf{0.1076} & \cellcolor{gray!20}+8.25\% \\
& H@20 & 0.0618 & 0.0540 & 0.0696 & 0.0646 & 0.0898 & 0.0838 & 0.1058 & 0.1184 & \underline{0.1256} & 0.0732 & 0.1106 & \textbf{0.1310} & \cellcolor{gray!20}+4.30\% \\
\midrule
\multirow{6}{*}{\textbf{Goodreads}}
& N@5  & 0.0071 & 0.0074 & 0.0254 & 0.0100 & \underline{0.0395} & 0.0381 & 0.0076 & 0.0387 & 0.0321 & 0.0107 & 0.0314 & \textbf{0.0420} & \cellcolor{gray!20}+6.33\% \\
& N@10 & 0.0109 & 0.0088 & 0.0293 & 0.0136 & 0.0430 & 0.0411 & 0.0114 & \underline{0.0436} & 0.0371 & 0.0121 & 0.0342 & \textbf{0.0470} & \cellcolor{gray!20}+7.80\% \\
& N@20 & 0.0152 & 0.0113 & 0.0327 & 0.0175 & 0.0471 & 0.0446 & 0.0152 & \underline{0.0474} & 0.0414 & 0.0176 & 0.0377 & \textbf{0.0515} & \cellcolor{gray!20}+8.65\% \\
& H@5  & 0.0126 & 0.0108 & 0.0362 & 0.0150 & 0.0502 & \underline{0.0532} & 0.0116 & \underline{0.0532} & 0.0408 & 0.0136 & 0.0408 & \textbf{0.0544} & \cellcolor{gray!20}+2.26\% \\
& H@10 & 0.0242 & 0.0152 & 0.0482 & 0.0264 & 0.0612 & 0.0626 & 0.0236 & \underline{0.0682} & 0.0560 & 0.0187 & 0.0496 & \textbf{0.0700} & \cellcolor{gray!20}+2.64\% \\
& H@20 & 0.0412 & 0.0252 & 0.0618 & 0.0418 & 0.0776 & 0.0766 & 0.0386 & \underline{0.0830} & 0.0734 & 0.0358 & 0.0636 & \textbf{0.0878} & \cellcolor{gray!20}+5.78\% \\
\bottomrule
\end{tabular}%
}
\label{tab:main}
\end{table*}

\subsubsection{Baselines.}
To evaluate the effectiveness of our method, we conduct a comprehensive comparison against the following traditional and LLM-based baselines:
\begin{itemize}[leftmargin=*]
\item {\textbf{Caser}~\cite{tang2018caser}}:
A CNN-based model that captures sequential patterns using horizontal and vertical convolutional filters.
\item {\textbf{GRU4Rec}~\cite{hidasi2018gru4rec}}: 
An RNN-based model employing gated recurrent units (GRUs) to encode the user's interaction sequences.
\item {\textbf{SASRec}~\cite{kang2018sasrec}}:
A Transformer-based model that uses a causal self-attention mechanism to capture item dependencies.
\item \textbf{Caser*, GRU4Rec*, SASRec*}: These are variants of the traditional methods above, where item embeddings are initialized with pre-trained LLM representations instead of being randomly initialized. This allows us to assess the benefit of semantic information from LLMs.
\item {\textbf{AlphaRec}~\cite{sheng2025language}}: 
An LLM-based collaborative filtering method using LLM-generated embeddings as a replacement for ID-based embeddings.
\item {\textbf{D$^3$}~\cite{bao2024decoding}}: 
An LLM fine-tuning method that introduces debiasing techniques during inference. Its fine-tuning process is similar to that of BIGRec~\cite{bao2025bi}. We disable its original ensemble design for a fair comparison.
\dhx{
\item {\textbf{S-DPO}~\cite{chen2024softmax}}: 
An LLM-based preference alignment method grounded in RL principles that extends the Direct Preference Optimization (DPO) algorithm to multi-negative scenarios following an SFT phase.
\item {\textbf{CoT-RL}~\cite{zhang2025reinforced}}: An LLM-based method that directly optimizing CoT generation via RL, similar to DeepSeek-R1-zero~\cite{guo2025deepseek}.
\item {\textbf{LatentR$^3$}~\cite{zhang2025reinforced}}: An LLM-based method that performs latent reasoning by generating compact latent thought tokens via SFT followed by RL, without relying on explicit CoT data.
}

\end{itemize}

\subsubsection{Evaluation Protocols.} 
We evaluate all methods on the next-item prediction task. Following standard practice~\cite{kang2018sasrec,bao2025bi,yang2023generic}, for each test interaction, the model predicts the next item based on the user's preceding history, including interactions from the test phase. 
Performance is measured using Normalized Discounted Cumulative Gain (NDCG@N) and Hit Ratio (HR@N) at 
$N \in \{5,10,20\}$~\cite{bao2025bi, rajput2023recommender}.
All methods are evaluated using the all-ranking protocol~\cite{krichene2020sampled}, where models rank all items. 
Due to the computational cost of generation, final evaluation is conducted on 5,000 instances randomly sampled from the test set. For LLM-based models, ranked lists are generated via beam search with a beam width of 20. Inspired by prior work~\cite{bao2024decoding}, we remove length penalty since it may be detrimental in this context.

\subsubsection{Implementation Details.}
For LLM-enhanced traditional methods (Caser*, GRU4Rec*, SASRec*) and LLM-based methods (RISER, AlphaRec, D$^3$, S-DPO, CoT-RL, LatentR$^3$), we use Qwen2-1.5B~\cite{team2024qwen2} as our backbone LLM.
Given the substantial computational cost of the reinforcement learning (RL) stage, RISER adopts a two-stage training methodology inspired by the common practice in LLMs—first SFT, followed by RL.  
In the first stage (SFT), the LLM is trained on the entire training dataset, denoted as $\mathcal{D}_{\text{SFT}}$, to align the model with the target domain and establish a foundational capability.  
In the second stage (RL), we randomly sample 10,000 instances from the original validation set to construct a new RL training set, $\mathcal{D}_{\text{RL}}$. Additionally, we select a disjoint set of 3,000 instances from the remaining data to serve as the validation set. This stage aims to unlock the model’s full potential by encouraging it to autonomously explore and refine its behavior.
To ensure a fair comparison, we subjected all baselines to the same two-stage training data protocol. After training on $\mathcal{D}_{\text{SFT}}$, each baseline was continually trained on $\mathcal{D}_{\text{RL}}$. This ensures that any performance advantage of RISER over these enhanced baselines is attributable to the RL algorithm itself, rather than mere exposure to additional data. Training hyperparameters of RISER and the baseline are detailed in Appendix~\ref{sec:implementation}.

\subsection{Main Results (RQ1)}
In this section, we analyze the top-N recommendation performance of RISER against baselines across three real-world datasets. The primary results, presented in Table~\ref{tab:main}, lead to the following key observations:
(1) RISER consistently outperforms all baselines across all metrics in both e-commerce and media domains, achieving substantial performance gains. For instance, it delivers average improvements of \dhx{over 25\% on Games and 7\% on Toys} in terms of both NDCG and HR when compared to the strongest baseline. These results underscore the effectiveness and robustness of RISER.
\dhx{
(2) RISER demonstrates a decisive advantage over reasoning-based approaches (i.e., CoT-RL and LatentR³). As observed, CoT-RL fails to deliver competitive performance, primarily due to the instability inherent in unsupervised reasoning generation. While LatentR$^3$ improves upon this by introducing latent tokens, it remains confined within the reasoning paradigm. In contrast, RISER completely bypasses these structural complexities. By leveraging RL for item space exploration, RISER avoids the pitfalls of compelling LLMs to reason over implicit behaviors, thereby establishing a more effective paradigm.
(3) RISER surpasses methods constrained by static supervision, including standard SFT (i.e., D$^3$) and preference alignment (i.e., S-DPO). We attribute this superiority to our RL-based approach, which optimizes for fully activating the potential of the LLMs. While D$^{3}$ is limited to imitating historical actions, and S-DPO focuses on aligning the model via a softmax ranking loss within a fixed data distribution, both remain confined to offline supervision. In contrast, RISER leverages active RL exploration to transcend these boundaries, which enables the discovery of more effective recommendation policies.}



\begin{table*}[t]
\centering
\caption{Ablation study on the Games dataset. We demonstrate the effect of incrementally adding each component to our baseline model RISER (SFT only). Avg. Improv. denotes the average cumulative gain over baseline at each step.}
\begin{tabular}{lcccccccc}
\toprule
\multirow{2}{*}{\textbf{Method}} & \multicolumn{4}{c}{\textbf{NDCG}} & \multicolumn{4}{c}{\textbf{HR}} \\
\cmidrule(lr){2-5} \cmidrule(lr){6-9}
& \textbf{@5} & \textbf{@10} & \textbf{@20} & \textbf{Avg. Improv.} & \textbf{@5} & \textbf{@10} & \textbf{@20} & \textbf{Avg. Improv.} \\
\midrule
Baseline & 0.0374 & 0.0444 & 0.0510 & - & 0.0496 & 0.0714 & 0.0976 & - \\
\midrule
+ GRPO & 0.0470 & 0.0539 & 0.0622 & +23.01\% & 0.0606 & 0.0822 & 0.1152 & +18.45\% \\
+ Length Penalty Removal & 0.0473 & 0.0540 & 0.0624 & +23.48\% & 0.0612 & 0.0820 & 0.1154 & +18.83\% \\
+ Oversampling \& De-duplication & 0.0496 & 0.0561 & 0.0631 & +27.57\% & 0.0652 & 0.0854 & 0.1138 & +22.55\% \\
+ KL-Cov & 0.0503 & 0.0573 & 0.0648 & +30.20\% & 0.0650 & 0.0868 & 0.1166 & +24.03\% \\
+ Certainty-Aware Loss Mask & 0.0507 & 0.0582 & 0.0655 & +31.69\% & 0.0664 & 0.0896 & 0.1188 & +27.03\% \\
+ SimPO & 0.0526 & 0.0598 & 0.0677 & +36.03\% & 0.0680 & 0.0902 & 0.1214 & +29.27\% \\
\bottomrule
\end{tabular}
\label{tab:ablation_study_final}
\end{table*}

\vspace{-0.2cm}
\subsection{In-depth Analysis}
In this subsection, we conduct an in-depth analysis of RISER.
We first conduct a comprehensive ablation study on its core components (RQ2). 
Next, we delve into its training stability, including an analysis of its performance across popular and unpopular items, and its entropy control (RQ3).
Third, we evaluate its scalability to assess how performance improves with increasing amounts of RL training data (RQ4).

\subsubsection{\textbf{Ablation Study (RQ2)}}
To quantify the contribution of each core component of RISER, we conduct an ablation study on the Games dataset. We begin with a baseline model trained only with SFT and incrementally add each component of RISER to observe its impact. The results are presented in Table~\ref{tab:ablation_study_final}, from which we observe that: (1) \textbf{RL provides a powerful but \dhx{volatile} foundation.} 
The initial application of a vanilla RL policy (GRPO) yields the largest performance leap (an average improvement of 23.01\% in NDCG and 18.45\% in HR over the SFT baseline). This indicates that the exploratory nature of RL is inherently more effective than SFT’s purely imitative learning.
\dhx{However, achieving this performance required extensive tuning—incorporating decoding constraints, aggressive KL penalties, and entropy loss—merely to enable training. Even with these adaptations, the vanilla RL policy remains fundamentally volatile, suffering from sudden performance drops and entropy collapse (as further analyzed in Sec. 4.3.2).} 
(2) \textbf{Stabilizing RL training is essential for sustained improvements.} 
Strategies and modules designed to enhance training stability (oversampling and de-duplication, adding KL-Cov and certainty-aware loss mask) consistently improve performance, raising the average NDCG gain from 23.01\% to 31.69\%. Notably, the inclusion of KL-Cov produces a clear boost, underscoring the importance of entropy control (later illustrated in Figure~\ref{fig:entropy}). These results confirm that mitigating RL’s inherent training instability is a prerequisite for consistent and substantial model gains.
(3) \textbf{Maximizing data utilization unlocks further performance potential.} With a stable policy in place, integrating SimPO delivers the second-largest improvement, elevating the average NDCG gain from 31.69\% to 36.03\%. 
This highlights that a significant portion of learning signal in vanilla GRPO resides in zero-advantage trajectories. By converting these wasted rollouts into useful learning signals, SimPO resolves a key data-efficiency bottleneck, enabling RISER to reach its full potential.

\begin{figure}[t]
    \setlength{\belowcaptionskip}{-0.5cm}
    \centering
    \scalebox{0.8}{
    \includegraphics[width=0.55\textwidth]
    {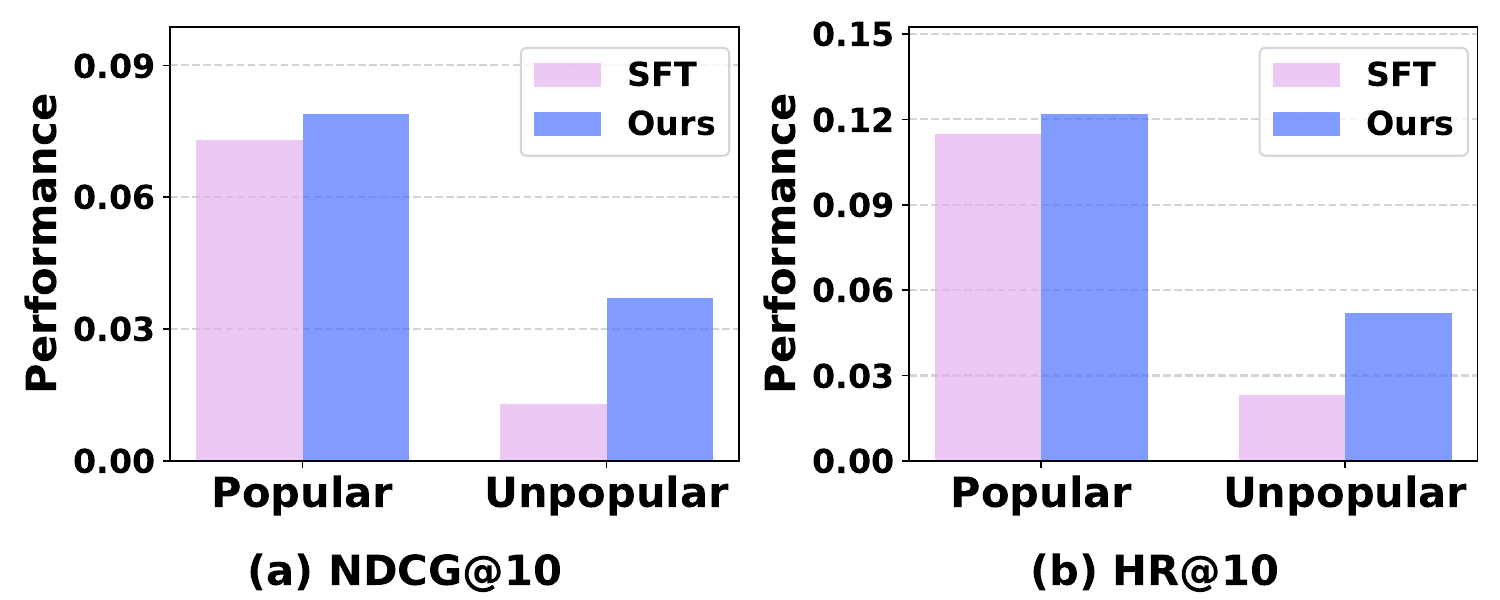}
    }
    \vspace{-0.2cm}
    \caption{
    Comparison of our method and SFT-trained method over popular and unpopular item groups on Games dataset, reporting NDCG@10 (Left) and HR@10 (Right).
    }
    \label{fig:pop}
    \vspace{0.2cm}
\end{figure}

\subsubsection{\textbf{Effect on Training Stabilization (RQ3)}}
To assess RISER’s impact on training stability, we conduct a two-part analysis. First, we compare performance on popular and unpopular items, since stabilizing the training process is crucial for preventing premature policy collapse toward popular items and thereby enhancing long-tail recommendation. Second, we analyze the entropy dynamics throughout training together with the corresponding HR on the test set, illustrating how RISER effectively regulates entropy and how such regulation translates into measurable performance gains.

\begin{itemize}[leftmargin=*]
\item{\textbf{Performance over Popular and Unpopular Items}}
To analyze the effectiveness of RISER on mitigating popularity bias, we partition items into popular (top 20\% by frequency in training set $\mathcal{D}_{\text{SFT}}$ and $\mathcal{D}_{\text{RL}}$ ) and unpopular (bottom 80\%) groups. We then compare RISER with the SFT baseline. 
As shown in Figure~\ref{fig:pop}, RISER achieves substantial improvements on the unpopular group, boosting HR@10 by 126.1\% and NDCG@10 by 184.6\%.
These results indicate that RISER effectively mitigates popularity bias by avoiding premature policy collapse onto popular items and successfully exploring the long-tail space. 
The improvement can be attributed to four key designs in RISER: 
(i) SFT models inherently exhibit a strong bias toward popular items. In contrast, RISER’s reinforcement learning objective assigns rewards to hard-to-retrieve unpopular items while penalizing popular items that are prone to errors, thereby creating a strong incentive to explore the long-tail items.  
(ii) SimPO enhances exploration efficiency by learning from zero-advantage rollouts—a scenario that frequently arises when sampling unpopular items.  
(iii) The oversampling and de-duplication strategy directly diversifies the training batch. By preventing the policy from repeatedly generating  the same few frequently generated items, it ensures broader coverage of the tail distribution.

\begin{figure}[t]
    \setlength{\belowcaptionskip}{-0.5cm}
    \centering
    \scalebox{0.85}{
    \includegraphics[width=0.55\textwidth]
    {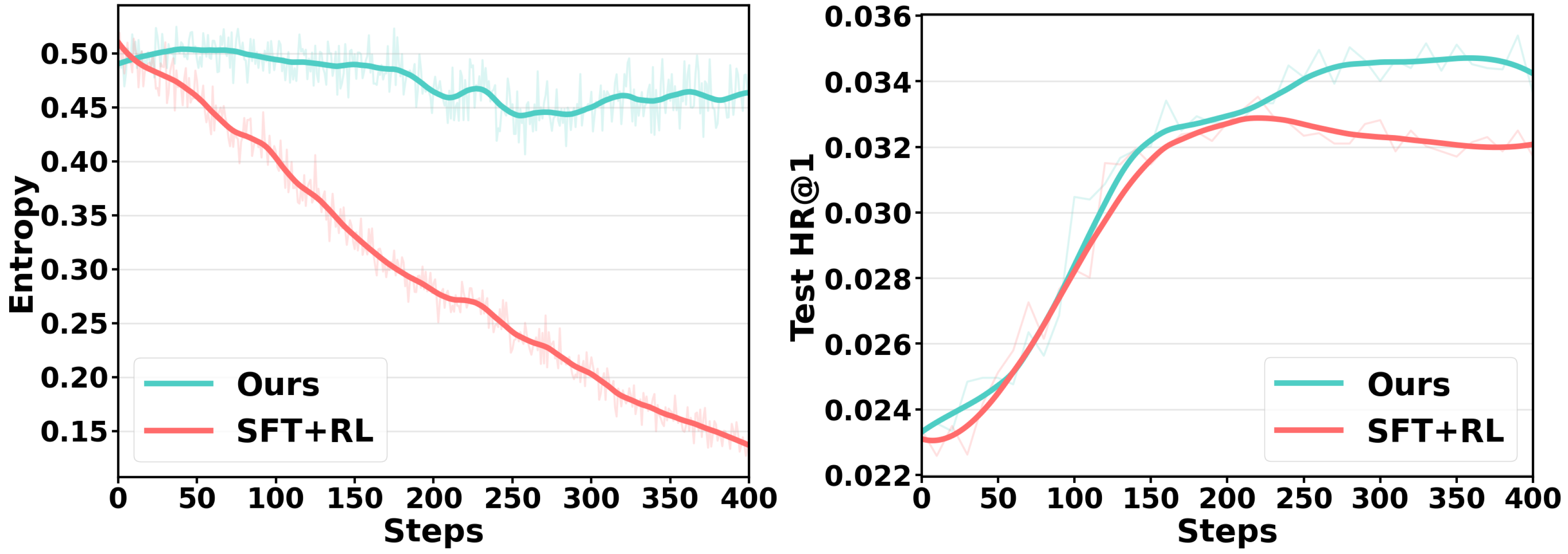}
    }
    \vspace{-0.5cm}
    \caption{
    Comparison of policy entropy (Left) and test set HR@1 (Right) for RISER and the SFT+RL baseline over the initial 400 training steps on the Games dataset.
    }
    \label{fig:entropy}
    \vspace{0.2cm}
\end{figure}

\begin{figure*}[t]
    \setlength{\belowcaptionskip}{-0.3cm}
    \centering
    \begin{subfigure}[t]{0.32\textwidth}
        \centering
        \includegraphics[width=\linewidth]{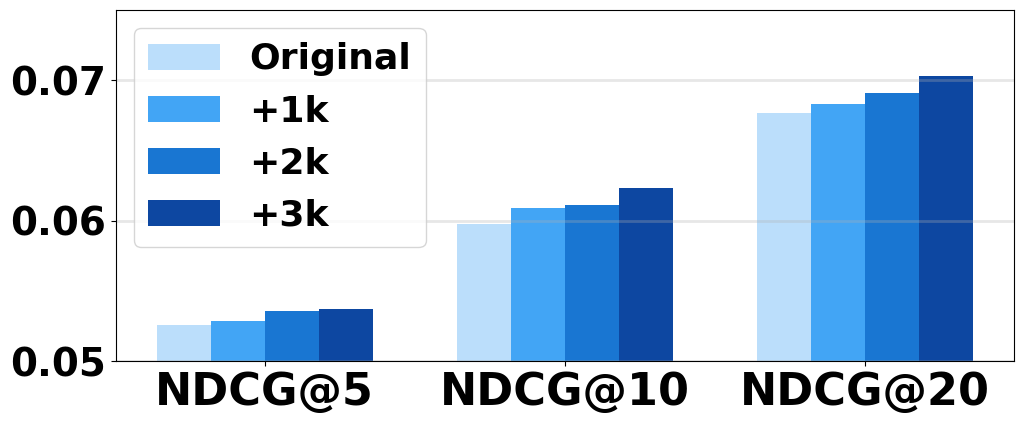}
        \end{subfigure}
    \hfill
    \begin{subfigure}[t]{0.32\textwidth}
        \centering
        \includegraphics[width=\linewidth]{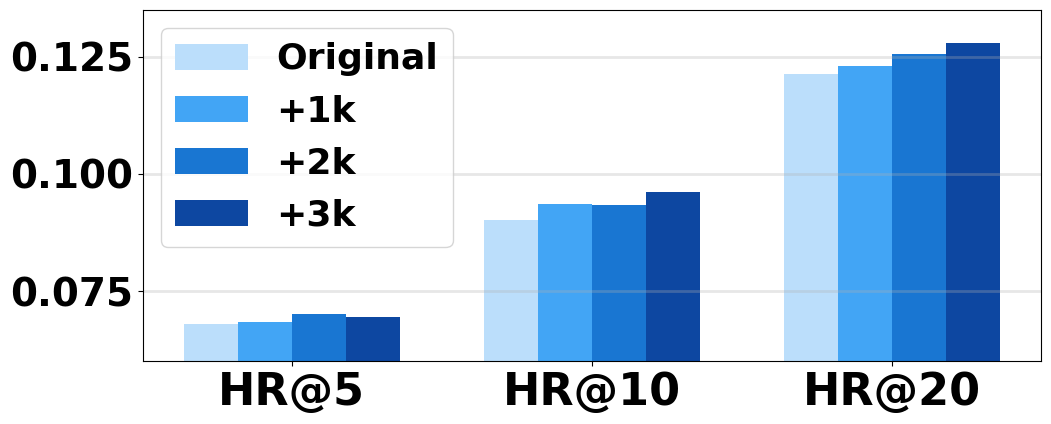}
    \end{subfigure}
    \hfill
    \begin{subfigure}[t]{0.32\textwidth}
        \centering
        \includegraphics[width=\linewidth]{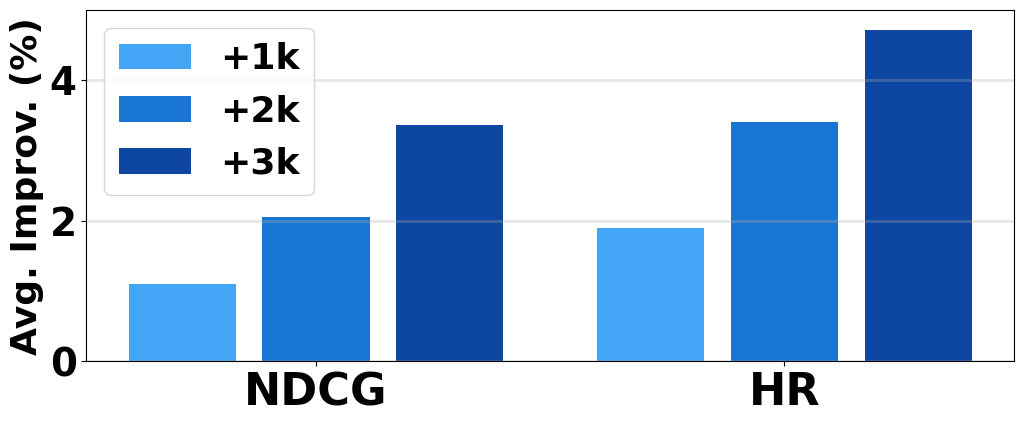}
    \end{subfigure}
    \caption{Scalability of RISER on the Games dataset when its RL training stage is augmented with 1k, 2k, and 3k additional ``hard positive'' samples. The plots show the resulting performance on NDCG@{5, 10, 20} (Left) and HR@{5, 10, 20} (Middle), and the relative improvement over the baseline with 0k additional data (Right).}
    \label{fig:scale}
\end{figure*}

\item{\textbf{Performance of Entropy Control}}
To investigate RISER's effectiveness in enhancing training stability, we compare its entropy dynamics and corresponding test set performance (HR@1) against those of a vanilla RL baseline \dhx{with extensive tuning specified in Sec. 4.3.1} throughout the training process. As shown in Figure~\ref{fig:entropy}\footnote{Full results are available in Appendix~\ref{sec:full_ent}.}, the vanilla RL baseline suffers from a rapid entropy collapse, an evident indicator of training instability. Consequently, its HR@1 is highly volatile and may even decline with additional training steps. In contrast, RISER maintains stable policy entropy, which facilitates consistent growth in HR@1, allowing it to ultimately surpass the baseline and converge at a higher level. 
This stability stems from three key components in RISER’s design: (i) the oversampling and de-duplication strategy provides a foundation of diverse explorations; (ii) the certainty-aware loss mask prevents over updates by downgrading gradients on predictable, non-branching tokens, thus stabilizing the policy distribution; and (iii) The KL-Cov regularizer complements this by selectively penalizing high-confidence, high-advantage outlier tokens, which tempers the disproportionate gradient updates that distinguishing tokens often receive due to batch imbalance and ensures a smoother policy update.


    
    
    

\end{itemize}
\subsubsection{\textbf{Performance with Increasing Training Data (RQ4)}}
We next evaluate RISER’s data scalability by assessing its ability to leverage additional data for further performance gains. To this end, we construct a pool of hard positive data from the Games training set $\mathcal{D}_{\text{SFT}}$. An item is included in this pool if it is ranked between positions 6 and 20 (i.e., a hit in HR@20 but not in HR@5) by the SFT-trained RISER. These items are predicted as relevant yet fail to achieve high ranks, representing clear opportunities for improvement. From this pool, we randomly sample and add 1k, 2k, and 3k instances to the RL training set $\mathcal{D}_{\text{RL}}$.
As shown in Figure~\ref{fig:scale}, RISER’s performance scales positively with the amount of training data. With 3k additional data, average NDCG improves by 3.37\% and average HR by 4.71\%. 
This demonstrates RISER's scalability and its ability to effectively learn from challenging data. 
The improvement stems from two factors: (i) RISER’s efficient learning design, extracts valuable signals from every exploration step, all newly encountered information is utilized effectively; and (ii) RISER's stabilization designs provide a robust learning environment, which is crucial for the model to reliably integrate the new information discovered through exploration without degrading its overall performance.

\section{Related Work}


\noindent $\bullet$ \textbf{LLM-based Recommendation.}
Recent years have witnessed the integration of LLMs into recommender systems, leveraging their strong semantic comprehension and generative capabilities~\cite{bao2023tallrec, rajput2023recommender}. 
LLM-based recommendation reframes recommendation as a natural language understanding and generation task. 
Existing LLM-based recommendation approaches can be broadly categorized into two paradigms: (1) directly leveraging existing LLMs for recommendation~\cite{gao2023chat, zhang2024agentcf}, and (2) fine-tuning LLMs for recommendation~\cite{bao2023tallrec, DBLP:journals/tois/ZhangXHZLW25}. In the first paradigm, researchers harness the powerful natural language understanding and generation capabilities of LLMs through techniques such as in-context learning~\cite{sun2024large, gao2023chat} or agent-based~\cite{zhang2024agentcf, wang2024recmind} frameworks to do recommendation. Although these direct-utilization methods achieve reasonable recommendation performance, the substantial discrepancy between the pretraining objectives of LLMs and recommendation limits LLM's recommendation ability. To bridge this gap, lots of work explores fine-tuning strategies to align LLMs with recommendation tasks~\cite{bao2023tallrec, DBLP:journals/tois/ZhangXHZLW25}.
Among fine-tuning approaches, SFT is the predominant method~\cite{msl, recranker, graphrec}. However, SFT is inherently limited to mimicking historical user–item interactions, which leads to intrinsic imitation biases—such as popularity bias—and related issues~\cite{item_bias}. To overcome the constraints of merely replicating past user–item interactions, several RL methods (e.g., GRPO~\cite{shao2024deepseekmath}) have been applied in LLM-based recommendation~\cite{zhang2025reinforced, lin2025recr1, zhang2025slow}. Nevertheless, these RL-based methods suffer from two major drawbacks: (1) during inference, they incur substantial latency overhead due to Long CoT; and (2) during training, they exhibit low sample efficiency and instability. In contrast, our proposed RISER addresses these challenges and demonstrates both efficient and effective recommendation performance.


\noindent $\bullet$ \textbf{RL for LLM.} RL has long been a cornerstone in aligning and enhancing LLMs. Its early prominence stemmed from Reinforcement Learning from Human Feedback (RLHF) \cite{rlhf2022}, which enabled LLMs to better reflect human preferences through reward modeling. However, the recent breakthroughs of models such as OpenAI’s o1 \cite{jaech2024openai} and DeepSeek-R1 \cite{guo2025deepseek} have catalyzed a paradigm shift—moving RL beyond preference alignment toward reasoning amplification. This surge in RL-driven reasoning has also influenced adjacent domains, including recommendation systems \cite{zhang2025reinforced, lin2025recr1, zhang2025slow}. Although GRPO has been introduced into the field of recommendation systems \cite{zhang2025reinforced, lin2025recr1, zhang2025slow}, these works introduce a critical research gap: sparse rewards lead to profound sample inefficiency, which severely limits RL's practical application in recommendation systems where fewer valid rewards are available.  In contrast to previous works like Rec-R1 \cite{lin2025recr1} and STREAM \cite{zhang2025slow}, which primarily rely on reward engineering, either by scaling the evaluation metric (e.g., using K=1000 in NDCG@K) or by crafting complex reward functions, our approach proposes a novel solution that overcomes the inherent limitations of such reward-engineering strategies.


\vspace{-0.2cm}
\section{Conclusion and Future Work}
In this paper, we first highlight the misalignment between Long CoT reasoning and practical SeqRec systems, stemming from its high inference latency and the \dhx{scarcity of explicit cognitive patterns in implicit behavioral data}. This observation motivates us to directly apply RL to LLM-based recommendation. However, we identify two critical challenges: severe sample inefficiency and training instability. To address these issues, we propose RISER, a novel end-to-end RL framework for LLM-based recommendation. RISER incorporates SimPO to effectively learn from failed rollouts, and introduces a suite of stability mechanisms—including oversampling with de-duplication, a certainty-aware masking strategy, and KL-Cov regularization—to ensure robust and stable training. Extensive experiments demonstrate that RISER significantly outperforms strong baselines, establishing a new and more effective paradigm for applying RL to LLM-based recommendation. In the future, we will proceed in two main directions: first, exploring more advanced token-level optimization techniques; and second, improving the exploration efficiency of RL for LLM-based recommendation to enhance overall training efficiency and scalability.

\bibliographystyle{ACM-Reference-Format}
\bibliography{main}


\appendix
\section{Prompt Structure}
\label{sec:prompt}
\begin{table}[h!]
    \centering
    \caption{The prompt structure used for SFT and RL, illustrated with an example from the Games dataset.}
    \label{tab:prompt_structure}
    \begin{tabular}{lp{0.7\columnwidth}} 
        \toprule
        \multicolumn{2}{l}{\textbf{Instruction Input}} \\
        \midrule
        \addlinespace
        \textit{Instruction:} & Given a list of video games the user recently enjoyed, please write a new video game that the user might also like. \\
        \addlinespace
        \textit{User Input:} & The user has played the following video games before: \\
                             & \quad "Crash Bandicoot" \\
                             & \quad "Final Fantasy XV - PlayStation 4" \\
                             & \quad "The Legend of Zelda: Breath of the Wild - Wii U" \\
                             & \quad "NieR: Automata - Playstation 4" \\
                             & \quad "Dragon Quest Builders - PlayStation 4" \\
        \addlinespace
        \textit{Response:}   & (Model begins generation here) \\
        \midrule
        \multicolumn{2}{l}{\textbf{Instruction Output}} \\
        \midrule
        \addlinespace
             & "Persona 5 - SteelBook Edition - PlayStation 4" \\
        \bottomrule
    \end{tabular}
\end{table}

\section{Dataset Statistics and Data Preprocessing}

\label{sec:dataset_statistics}
\vspace{-10pt}
\begin{table}[H]
\centering
\caption{Dataset statistics.}
\vspace{-5pt}
\label{tab:dataset_stats}
\begin{tabular}{lrrrr}
\toprule
\textbf{Dataset} & \textbf{\# Train} & \textbf{\# Valid} & \textbf{\# Test} &\textbf{\# Items} \\
\midrule
Games      & 201,612    &25,201& 25,202 & 11,037\\
Toys       & 112,754    &14,094 & 14,095 &  11,252\\
Goodreads  & 106,131    & 13,266 &13,267 & 10,583 \\
\bottomrule
\end{tabular}
\end{table}

\vspace{-5pt}
To balance the significant computational cost of training LLMs while ensuring dataset validity, we employ a dynamic temporal partitioning strategy. The core of this strategy is to iteratively process the data within a sliding time window. For our games dataset for example, we first define an initial time window of user activity from October 2017 to October 2018. After applying 5-core filtering to this data, we check the number of unique games. If this count is below 10,000, we expand the window by shifting its start time backward by 3 months to July 2017, while the end time remains fixed at October 2018. This adjustment process is repeated until the processed dataset contains more than 10,000 unique games.

\section{Implementation Details}
\label{sec:implementation}
RISER adopts a two-stage training methodology. For the SFT stage, we use the AdamW optimizer~\cite{loshchilovdecoupled}. We search for the optimal learning rate in $\{3e^{-3,-4,-5}\}$with early stopping (patience=1).
For the RL stage, we tune several key hyperparameters: the mask decay coefficient $d$ is selected from \{0.5, 0.6, 0.7, 0.8\} (Eq.~\ref{eq:certainty_mask}); the rollout temperature $\tau$ from \{0.6, 0.8, 1.0, 1.2\}; the SimPO hyperparameter $\beta_{Simpo}$ from \{5, 10\}; and the gamma-beta ratio $\gamma/\beta$ from \{0.7, 1.0\} (Eq.~\ref{eq:simpo_loss_final}).  Other parameters are fixed: The learning rate is 1e-6. The KL-Cov ratio $k$ is 5e-3 and the KL coefficient $\beta_{\text{KL}}$ is 1.0 (Eq.~\ref{eq:kl_cov}). For policy rollout, we generate an initial $m=20$ sequences, which are then de-duplicated to yield up to $n=16$ trajectories for training (Eq.~\ref{eq:oversampling}). Experiments are conduct on 2 NVIDIA A100 GPUs. 

For traditional baselines (Caser, GRU4Rec, SASRec), we use the Adam optimizer~\cite{kingma2014adam} with BCE loss, where negative items are randomly sampled. We search for the optimal learning rate in \{$10^{-2, -3,-4}$\} and weight decay in \{$10^{-3, \dots, -7}$\}. The embedding dim is 64 and the batch size is 256. For Caser$^*$, we project its LLM-initialization dimension to 256 for effective computation. 
\dhx{For the remaining methods, we follow the default settings reported in their respective papers.}

\vspace{-1.5pt}
\section{Full Entropy Control Results}
\label{sec:full_ent}
\begin{figure}[H]
    \centering
    \scalebox{0.8}{
    \includegraphics[width=0.55\textwidth]
    {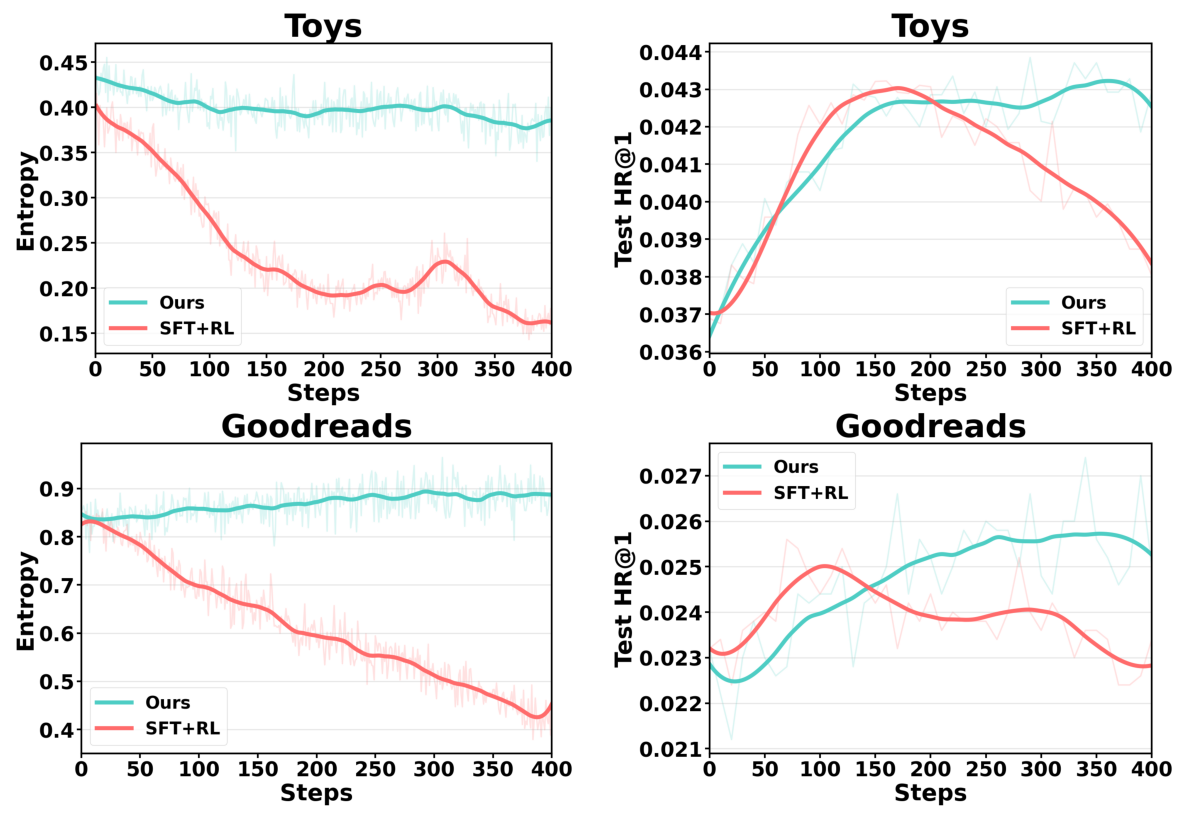}
    }\vspace{-0.5pt}
    \caption{Comparison of policy entropy and test set HR@1 for RISER and the SFT+RL baseline over the initial 400 training steps on the Toys and Goodreads datasets.
    }
    \label{fig:ent_full}
\end{figure}

Figure~\ref{fig:ent_full} provides the complete results of entropy control experiments discussed in Sec. 4.3.2, consistent with the conclusions drawn in the main text.

\section{RISER Algorithm}
\label{sec:algo}
The complete RISER training process is detailed in Algorithm~\ref{alg:riser}.

\begin{algorithm}[H]
\caption{The RISER Training Algorithm}
\label{alg:riser}
\begin{algorithmic}[1]
    \Statex \textbf{Phase 1: SFT}
    \State Train policy $\pi_\theta$ on $\mathcal{D}_{\text{SFT}}$ using $\mathcal{L}_{\text{SFT}}$ (Eq.~\ref{eq:sft_objective}).
    \Statex
    \Statex \textbf{Phase 2: RL}
    \State \textbf{Initialize:} Initialize current policy $\pi_\theta$ and sampling policy $\pi_{\theta_{\text{old}}}$ with SFT weights.
    \For{each training iteration}
        \State Sample a batch of prompts $\mathcal{B}$ from $\mathcal{D}_{\text{RL}}$.
        \State Generate rollouts $\mathcal{O}_{\text{Rollout}}$ from $\mathcal{B}$ via $\pi_{\theta_{\text{old}}}$ with Oversampling \& De-duplication (Eq.~\ref{eq:oversampling}).
        \State Partition $\mathcal{O}_{\text{Rollout}}$ into successful set $\mathcal{O}_{\text{Succ}}$ and failed set $\mathcal{O}_{\text{Fail}}$.
        \State Compute covariance scores $\text{Cov}(t)$ for all tokens in $\mathcal{O}_{\text{Succ}}$ (Eq.~\ref{eq:kl_cov}).
        \State Identify the set of outlier tokens $\mathcal{T}_{\text{outliers}}$ from top-$k$ covariance scores.
        \State Compute the modified GRPO loss $\mathcal{L}'_{\text{GRPO}}$ on $\mathcal{O}_{\text{Succ}}$ (Eq.~\ref{eq:grpo_prime_loss}), applying the certainty-aware mask (Eq.~\ref{eq:certainty_mask}).
        \State Construct the set of preference pairs $\mathcal{P}_{\text{Pref}}$ from $\mathcal{O}_{\text{Fail}}$.
        \State Compute the SimPO loss $\mathcal{L}_{\text{SimPO}}$ on $\mathcal{P}_{\text{Pref}}$ (Eq.~\ref{eq:simpo_loss_final}), applying the certainty-aware mask (Eq.~\ref{eq:certainty_mask}).
        \State Combine the losses: $\mathcal{L}_{\text{RISER}} \leftarrow  w \cdot \mathcal{L}_{\text{SimPO}} + \mathcal{L}'_{\text{GRPO}}$.
        \State Update the current policy parameters $\theta$ using the gradient $\nabla_\theta \mathcal{L}_{\text{RISER}}$.
        \State Update the sampling policy for the next iteration: $\pi_{\theta_{\text{old}}} \leftarrow \pi_\theta$.
    \EndFor
\end{algorithmic}
\end{algorithm}

\end{document}